\documentclass[prb,twocolumn,a4paper,showpacs,floatfix,superscriptaddress]{revtex4}  

\usepackage{graphicx}
\usepackage{amsmath}
\usepackage{amssymb}
\usepackage{amsthm}
\usepackage{color}

\graphicspath{{./IMAGES/}}

\begin{document}
\title{Massless Dirac fermions in two dimensions: Confinement in nonuniform magnetic fields}

\author{C. A. Downing}
\email[]{downing@ipcms.unistra.fr}
\affiliation{Institut de Physique et Chimie des Mat\'{e}riaux de Strasbourg, Universit\'{e} de Strasbourg, CNRS UMR 7504, F-67034 Strasbourg, France}

\author{M. E. Portnoi}
\email[]{m.e.portnoi@exeter.ac.uk} 
\affiliation{School of Physics,
University of Exeter, Stocker Road, Exeter EX4 4QL, United Kingdom}
\affiliation{International Institute of Physics, Universidade Federal do Rio Grande do Norte, 59012-970 Natal - RN, Brazil}

\date{\today}

\begin{abstract}
We show how it is possible to trap two-dimensional massless Dirac fermions in spatially inhomogeneous magnetic fields, as long as the formed magnetic quantum dot (or ring) is of a slowly-decaying nature. It is found that a modulation of the depth of the magnetic quantum dot leads to successive confinement-deconfinement transitions of vortex-like states with a certain angular momentum, until a regime is reached where only states with one sign of angular momentum are supported. We illustrate these characteristics with both exact solutions and a hitherto unknown quasi-exactly solvable model utilizing confluent Heun functions.
\end{abstract}

\maketitle

\section{\label{intro}Introduction}

Inhomogeneous magnetic fields continue to play an important role in modern physics, from the historic Stern-Gerlach experiment\cite{Gerlach} of 1922 to the post-World War II achievements in magnetic confinement of plasmas in tokamaks\cite{Sakharov} and the more recent magnetic levitation of macroscopic objects.\cite{Berry}

With the rise of the two-dimensional (2D) Dirac materials such as graphene, whose electrons behave like massless Dirac fermions, the influence of magnetic fields has been pivotal to  research into a range of fundamental physics,\cite{Goerbig} including relativistic Landau levels,\cite{McClure, Haldane, Zheng} Fock-Darwin states,\cite{Chen} integer\cite{Novoselov} and fractional\cite{Bolotin} quantum Hall effects, Hofstadter butterflies\cite{Hunt} and quantum spin Hall states.\cite{Young}

An important feature of Dirac fermions is the complete absence of backscattering,\cite{Katsnelson} leading to a great difficulty in confining electrons electrostatically.\cite{Rozhkov, Stone} Therefore, much effort has been expounded on considerations of magnetic traps.\cite{LeeReview, NogaretReview, arxiv} One-dimensional magnetic confinement has been shown to be key for snake states\cite{Snake1, Snake2, Snake3} and many inhomogeneous field profiles have been treated.\cite{Park1, Semi, Ghosh1, Kuru, Ghosh2, Tan, Milpas}  Zero-dimensional confinement in perpendicular magnetic fields has also been treated extensively,\cite{Martino, Masir, Wang, Hausler, Wang2, Lee, Slobodeniuk, Roy} with magnetic antidots\cite{Martino} and antirings\cite{Wang, Lee} (where the magnetic field is zero inside the dot and ring respectively) being shown to confine electrons. However, magnetic quantum dots (where the magnetic field is nonzero inside the dot) has been shown for a square-well magnetic field\cite{Masir} to not support bound states. In fact, as long as the magnetic trap decays at infinity slowly, bound states are indeed possible in both magnetic dots and rings, as we show in this work. 

Here we discuss several examples of magnetic profiles which show confinement in magnetic quantum dots and rings is achievable. Notably, we study smooth magnetic fields, which are both spatially inhomogeneous and regular at the origin, rather than the well-known square well models. In doing so, we make use of both exact solutions and a quasi-exactly solvable\cite{Turbiner, HartmannPortnoi} (QES) model, which most clearly display the underlying physics. Furthermore, studies of inhomogeneous electric fields\cite{Downing2, Downing3} have already been shown to be important in Dirac materials as compared to electrostatic square well models. We find here that a characteristic of magnetic quantum dots with decreasing field strength is the removal one-by-one of quantum states with diminishing angular momentum, until a plateau is reached where only bound states with negative angular momentum exist. 

Experimentally, nonuniform magnetic fields\cite{Katsumoto} can be created by various means, including deposition of ferromagnetic microstructures\cite{Ye} or superconducting stripes on top of the 2D electron gas,\cite{Carmona} or by curving the membrane.\cite{Foden} Indeed, a recent experiment\cite{Lindvall} has successfully studied weak localization in graphene in inhomogeneous magnetic fields (created by a thin film of type-II superconducting niobium in close proximity to the graphene layer). To obtain circularly symmetric magnetic fields like those considered in this work, one can utilize the field generated from a circular loop of current, when the loop has a large radius compared to the electronic sample so that the field is necessarily of a slowly-decaying nature.

\begin{figure}[tbp]
 \includegraphics[width=0.45\textwidth]{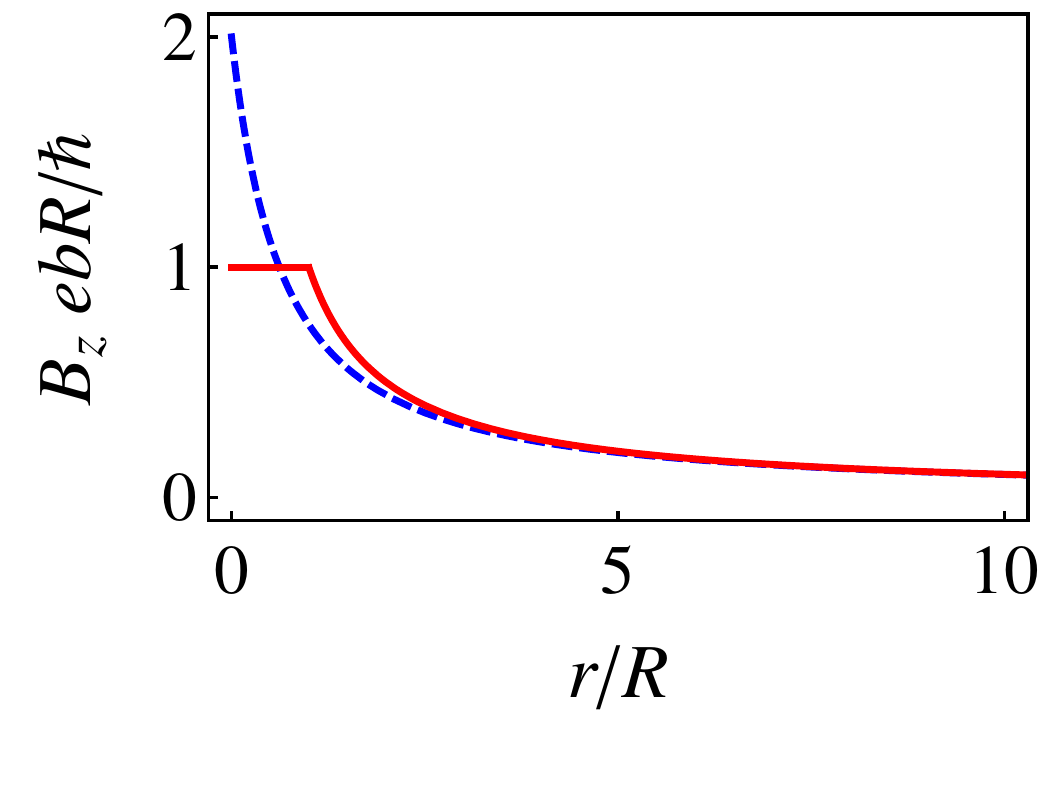}
 \caption{(Color online) Plots of the magnetic traps considered: the regularized magnetic quantum dot (solid red line) and the smooth magnetic quantum dot (dashed blue line).}
 \label{fig1}
\end{figure}

The single particle Hamiltonian describing the 2D excitations in graphene and other such Dirac-Weyl materials in a magnetic field $\boldsymbol{B} = \nabla \times \boldsymbol{A}$ can be written
\begin{equation}
\label{intro1}
\hat H = v_{\mathrm F} \boldsymbol \sigma \cdot \left( \boldsymbol{\hat p} + e \boldsymbol{A} \right),
\end{equation}
where $v_F$ is the Fermi velocity, $\boldsymbol \sigma = \left( \sigma_x, \sigma_y \right)$ are Pauli's spin matrices and $\boldsymbol A$ is a magnetic vector potential. Acting with this Hamiltonian on a wavefunction of the form 
\begin{equation}
\label{intro2}
\Psi(r,\theta) = \frac{e^{im\theta} }{\sqrt{2\pi}} \left(
 \begin{array}{c}
\chi_A(r) \\ ie^{i\theta}\chi_B(r)
 \end{array}
\right), \quad m=0,\pm1, \pm2, ...
\end{equation}
leads to the following coupled equations
\begin{subequations}
\label{intro3}
 \begin{align}
  \left(\partial_r + \tfrac{m+1}{r} + \tfrac{e}{\hbar} A_{\theta} \right) \chi_B &= \varepsilon\chi_A, \label{coupledupper} \\
  \left(-\partial_r + \tfrac{m}{r} + \tfrac{e}{\hbar} A_{\theta} \right) \chi_A &= \varepsilon\chi_B, \label{coupledlower}
 \end{align}
\end{subequations}
where the eigenvalue $E= \hbar v_F \varepsilon$, and the perpendicular magnetic field $B_z = r^{-1} \partial_r (r A_{\theta} (r))$ enters via the angular component of the vector potential. The wavefunction~\eqref{intro2} is an eigenfunction of the total angular
momentum operator $J_z = -i \hbar \partial_{\theta} + \hbar \sigma_z /2$, such that $J_z \Psi = (m+1/2) \Psi$. The compact equations~\eqref{intro3} form the basis of this work. In Sec.~\ref{sec2} we treat a slowly-decaying, regularized magnetic quantum dot, and go on to consider a spatially smooth magnetic quantum dot profile in Sec.~\ref{sec3}, please see Fig.~\ref{fig1} for sketches of these magnetic traps. We draw some conclusions in Sec.~\ref{conc} and detail corresponding results for magnetic quantum rings in Appendix~\ref{sec5}.

\section{\label{sec2}Dirac electron in a regularized magnetic quantum dot}

Let us consider a magnetic quantum dot, defined by
\begin{equation}
\label{eq30}
	B_z (r) = \frac{\hbar}{e} \frac{1}{b} \begin{cases} R^{-1}, \quad r \le R, \quad \text{(region I)} \\
r^{-1}, \quad r > R. \quad \text{(region II)} \end{cases}
\end{equation}
where we have introduced the length scale $R$ to regularize the field as $r \to 0$. This magnetic field profile is sketched in Fig.~\ref{fig1} as the solid red line. $R$ is in immediate competition with $b$, the parameter which effectively describes the magnitude of the field, via the key ratio $R/b$. The solutions of Eqs.~\eqref{intro3} in region I are the well known relativistic Landau level wavefunctions,\cite{Zheng} which can be given in terms of Kummer's function $F(a, b, z)$ as
\begin{equation}
\label{eq31}
	\chi_B^{I} = \tfrac{c_{I}}{b} \times (r/b)^{|m+1|} e^{-\tfrac{r^2}{4bR}} F(a_{I}, b_{I}, \tfrac{r^2}{2bR}),
\end{equation}
\begin{equation*}
\label{eq32}
	a_{I} = \tfrac{1}{2} \left( 1 + m + |1+m| - \varepsilon^2 b R \right), \quad b_{I} = 1 + |1+m|,
\end{equation*}
with the normalization constant $c_{I}$. The upper wavefunction component $\chi_A$ can be easily found from Eq.~\eqref{coupledupper}. The wavefunctions in region II can be found by reducing the system of equations~\eqref{intro3} to a single Schr\"{o}dinger equation for wavefunction component $\chi_B$ only. In this case one finds a formal identification with the 2D hydrogen atom,\cite{Parfitt} leading to 
\begin{equation}
\label{eq33}
	\chi_B^{II} = \tfrac{c_{II}}{b} \times (r/b)^{|m+1|} e^{-\tfrac{\kappa r}{b}} U(a_{II}, b_{II}, \tfrac{2 \kappa r}{b}),
\end{equation}
\begin{equation*}
\label{eq34}
	a_{II} = \tfrac{1}{2} + |1+m| + \tfrac{2m+1}{2 \kappa}, \quad b_{II} = 1 + 2|1+m|,
\end{equation*}
except here we choose instead the second linearly independent solution to Kummer's equation $U(a, b, z)$,\cite{Gradshteyn} in order to have a square-integrable wavefunction at infinity. In Eq.~\eqref{eq33} we have introduced $\kappa = \sqrt{1 - \varepsilon^2 b^2} > 0$ and $c_{II}$ is some constant.

Enforcing both wavefunction components to be continuous across the interface at $r=R$, one obtains the matching constant 
\begin{equation}
\label{eq35}
	\frac{c_{II}}{c_{I}} = e^{\tfrac{R}{b}\left(\kappa - \tfrac{1}{4}\right)} \frac{F(a_{I}, b_{I}, \tfrac{R}{2b})}{U(a_{II}, b_{II}, \tfrac{2 \kappa R}{b})},
\end{equation}
and the following rich transcendental equation for energy quantization, to be solved by root-finding methods
\begin{multline}
\label{eq36}
	\frac{a_{I}}{b_{I}} \frac{F(a_{I}+1, b_{I}+1, \tfrac{R}{2b})}{F(a_{I}, b_{I}, \tfrac{R}{2b})} + 2 \kappa a_{II} \frac{U(a_{II}+1, b_{II}+1, \tfrac{2 \kappa R}{b})}{U(a_{II}, b_{II}, \tfrac{2 \kappa R}{b})} \\
 + \kappa - 1 = 0, 
\end{multline}
subject to the bound $|\varepsilon b | < 1$. Equation~\eqref{eq36} interpolates between two simple expressions in the limiting cases of (i) a constant field, when $R/b >> 1$;  and (ii) a singular field, in the regime $R/b << 1$:
\begin{subequations}
\label{eq37}
 \begin{align}
  \varepsilon_{n, m} (b R)^{1/2} = \pm \left( 1 + m + |1+m| + 2n\right)^{1/2}, \tfrac{R}{b} \gg 1, \label{37a} \\
  \varepsilon_{n, m} b = \pm \left\{ 1 - \left( \frac{1+2 m}{1+2n+2 |m+1|} \right)^2 \right\}^{\tfrac{1}{2}}, m \le -1, \tfrac{R}{b} \ll 1, \label{37b}
 \end{align}
\end{subequations}
where $n=0,1, 2...$ is a quantum number. Equation~\eqref{37a}, in the constant magnetic field limit, describes the celebrated relativistic Landau levels, with the $\pm$ entering due to the presence of both electron and hole excitations in the system. These Landau levels are highly degenerate, as can be seen from the contributions of both quantum numbers $n$ and $m$ respectively. Equivalently, the spectrum Eq.~\eqref{37a} may be rewritten in a form familiar from calculations in the Landau gauge, $\varepsilon_{l} (b R)^{1/2} = \pm \left( 2 l \right)^{1/2}$, where $l$ is a nonnegative integer. These bound states are associated with a localization length $\zeta$, the lower bound of which can be estimated from the exponent in Eq.~\eqref{eq31} to be $\zeta \sim 2 \sqrt{b R}$.

In the opposite limit, Eq.~\eqref{37b} demonstrates that the spectrum is still dependent on both quantum numbers $n$ and $m$. There are highly nodal states near the maximal energy bounds $\varepsilon_{n>>1, m} b \simeq \pm 1$, and a reservoir of high-$|m|$ states near the Dirac point energy $\varepsilon_{n, m<<-1} b \simeq 0$, as well as many states in between. These bound states are characterized by a localization length $\zeta$ which is strongly dependent on the position of the energy level, and its lower bound may be estimated from the exponent in Eq.~\eqref{eq33} such that $\zeta \sim b / \kappa$. Whilst low-lying (and indeed zero energy) states have $\zeta \sim b$, high-lying states with $|\varepsilon b| \to 1$ are distinguishable by their large localization length $\zeta \to \infty$.  

\begin{figure}[tbp]
 \includegraphics[width=0.45\textwidth]{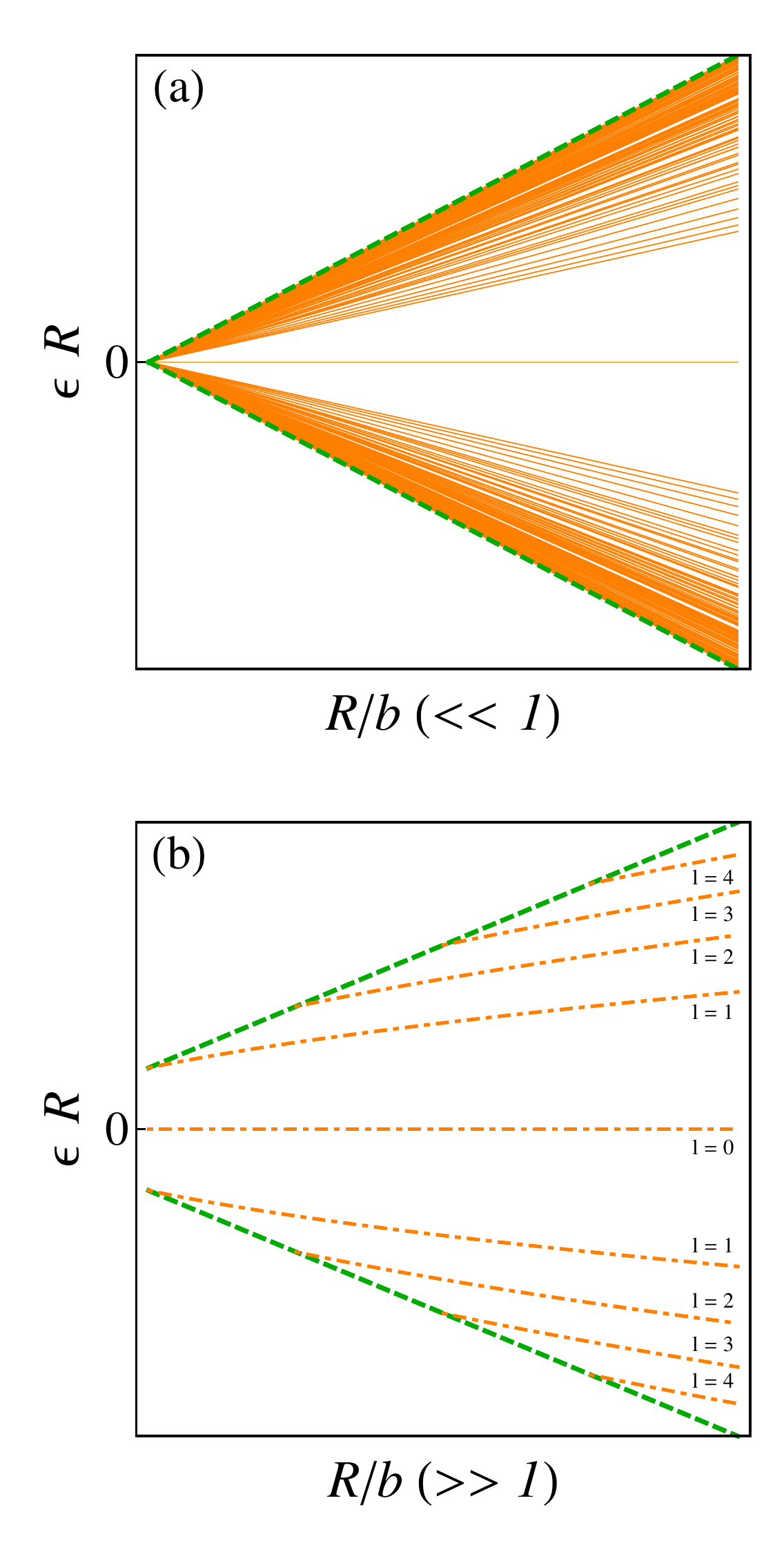}
 \caption{(Color online) Sketch of the energy levels $\varepsilon R$ as a function of the field parameter $R/b$, where the thick, dashed green lines $\varepsilon R = \pm R/b$ demarcate the region where bound states may form. In (a), $R/b \ll 1$ and the energy levels of Eq.~\eqref{37a} (solid, thin orange lines) shown are limited to those with $n = (0,...10)$ and $m = (-1,..,-10)$ for clarity. In (b), $R/b \gg 1$ and the highly degenerate energy levels of Eq.~\eqref{37b} (dash-dot, thin orange lines) are assigned the number $l$, which encompasses a plethora of combinations of the quantum numbers $n$ and $m$.}
 \label{fig2}
\end{figure}

The energy levels of both of the limiting cases (Eq.~\eqref{eq37}) are plotted in Fig.~\ref{fig2}, where the thick, dashed green lines $\varepsilon = \pm 1/b$ define the boundaries between which bound states may form. The regime of $R/b \ll 1$ is sketched in Fig.~\ref{fig1}~(a), where the solid, orange lines represent the plethora of energy levels which comprise the fan diagram. In Fig.~\ref{fig1}~(b), the situation of $R/b \gg 1$ is sketched, where the dash-dot orange lines depict the highly degenerate Landau-like energy levels, which may be ascribed the quantum number $l$. Notably, in this regime and with increasing $R/b$, at critical values of the magnetic field parameter $(R/b)_c = 2 l$ higher Landau-like levels are sequentially established. This is because the energy levels become residents of the sector of allowable bound state energies $-1/b < \varepsilon < 1/b$.

Now, upon decreasing $R/b$ one makes a transition from the ($R/b \gg 1$) limit of Eq.~\eqref{37a} to the ($R/b \ll 1$) regime governed by Eq.~\eqref{37b}, where the positive sign of angular momentum is now excluded. This suggests that a modulation of the magnetic field strength will allow one to observe successive confinement-deconfinement transitions of positive angular momentum states as they disappear into the continuum one-by-one with changing magnetic field strength. Such a phenomena of bound states diving into the continuum has the superficial appearance of being a magnetic version of the famous atomic collapse \cite{Zeldovich, Downing}. There the relativistic atom is modeled with a massive Dirac Hamiltonian in an external Coulomb field, and bound states merge from the gapped region into the continuum at critical charge strengths.

Striking in Fig.~\ref{fig2} is the presence of pure zero-energy states $(\varepsilon = 0)$. Zero-energy state solutions of various Dirac equations are of great interest due to their importance for fractional charge, the quantum Hall effect, topology, localization and Majorana physics.\cite{Jackiw, HartmannRobinson} Here zero-modes arise from the decoupled first-order equations~\eqref{intro3} that are readily integrated to yield:
\begin{equation}
\label{eq123456}
\left( \begin{array}{c}
\chi_A \\ \chi_B
 \end{array} \right) =  \left(
 \begin{array}{c}
 c_A r^m e^{ \mathcal{A}(r)} \\  c_B r^{-(m+1)} e^{- \mathcal{A}(r)}
 \end{array}
\right), \quad \mathcal{A}(r) = \frac{e}{\hbar} \int^r A_{\theta}(r') \mathrm{d}r'
\end{equation}
where $c_{A, B}$ are integration constants. It follows for the specific magnetic field consider here, Eq.~\eqref{eq30}, that
\begin{equation}
\label{388}
\left( \begin{array}{c}
\chi_B^I \\ \chi_B^{II}
 \end{array} \right) \sim r^{-(m+1)} \left(
 \begin{array}{c}
 e^{-\tfrac{r^2}{4 b R}} \\ e^{\tfrac{3 R}{4 b} - \tfrac{r}{b}}
 \end{array}
\right), ~ \chi_A^I = \chi_A^{II} = 0, ~ m \le -1.
\end{equation}
Thus these degenerate ground states are chiral: the wavefunction is nonzero only on one sublattice site ($\chi_B$), and they are equally shared between electrons and holes. Notably, the restriction on the sign of angular momentum is maintained. In graphene, the analogous result for the second valley is obtained by interchanging the upper and lower wavefunction components.  

\section{\label{sec3}Dirac electron in a smooth magnetic well}

It is expedient to check the physics found in above model is maintained for a smoothly regularized magnetic quantum dot, shown in Fig.~\ref{fig1} as the dashed, blue line. Thus we are led to consider the field 
\begin{equation}
\label{eq20}
 B_z (r) = \frac{\hbar}{e} \frac{1}{b R} \frac{2+r/R}{(1+r/R)^2}.
\end{equation}
We seek the lower wavefunction component in the form
\begin{equation}
\label{eq21}
	\chi_B = \tfrac{c}{b} \times \xi^{|m+1|} e^{-\kappa \xi} w(\xi),
\end{equation}
with $c$ a normalization constant and we use the notation
\begin{equation}
\label{eq22}
	\xi = r/b, \quad \kappa = (1 - \varepsilon^2 b^2)^{1/2} > 0.
\end{equation}
Equation~\eqref{eq20} is reasonable choice, since we know the behavior of the function as $\xi \to 0$ should be $\chi_B \sim \xi^{|m+1|}$, with the exponential decrease characteristic of a bound state $(e^{-\kappa \xi})$ as $\xi \to \infty$. Substitution of Eq.~\eqref{eq21} into Eq.~\eqref{intro3} and elimination of $\chi_A$ yields the equation 
\begin{equation}
\label{eq23}
	w''(\xi) + \left( \tfrac{1 + 2 |m+1|}{\xi} - 2 \kappa \right) w'(\xi) 
		+  \left( \Omega - \tfrac{\Upsilon}{\xi}  \right) w(\xi)  = 0, 
\end{equation}
\begin{equation*}
\label{eq23333}
	\Omega = \frac{1+b/R+2 \xi b/R}{(1+\xi b/R)^2} - \frac{(1 + 2 m) b/R}{1 + \xi b/R}, \quad \Upsilon = \kappa + 2 \kappa |m+1|,
\end{equation*}
where the prime denotes differentiation with respect to $\xi$. It is natural to introduce the new independent variable $\zeta = 1 + \xi b/R = 1 + r/R$, and after we take the ansatz
\begin{equation}
\label{eq24}
	w(\zeta) = \zeta^{1 - R/b} g(\zeta),
\end{equation}
we obtain a form of the confluent Heun equation\cite{Ronveaux}
\begin{equation}
\label{eq25}
	g''(\zeta) + \left( \alpha + \tfrac{\beta+1}{\xi} + \tfrac{\gamma+1}{\xi-1} \right) g'(\zeta) +  \left( \tfrac{\mu}{\xi} + \tfrac{\nu}{\xi-1} \right) g(\zeta) = 0,
\end{equation}
\begin{multline*}
\label{eq26}
	\text{where} \quad \mu = \tfrac{1}{2} \left( \alpha - \beta - \gamma + \alpha \beta - \beta \gamma \right) - \eta, \\
		\nu =\tfrac{1}{2} \left( \alpha + \beta + \gamma + \alpha \gamma + \beta \gamma \right) + \delta + \eta.
\end{multline*}
Explicitly the parameters are found to be
\begin{multline}
	\alpha = -2 \kappa \tfrac{R}{b}, \quad \beta = 1- 2 \tfrac{R}{b}, \quad \gamma = 2|m+1|,  \\
    \delta = 2 \tfrac{R^2}{b^2} - (1 + 2 m) \tfrac{R}{b}, \quad \eta = \tfrac{1}{2} + \tfrac{R}{b} (1 + 2 m) - 2 \tfrac{R^2}{b^2} .
\end{multline}
The Frobenius solution to Eq.~\eqref{eq25} is computed as a power series expansion around the origin $\zeta = 0$, a regular singular point, with a radius of convergence $|\zeta| < 1$
\begin{equation}
\label{eq28}
	g(\zeta) = \sum_{n=0}^{\infty} v_n(\alpha, \beta, \gamma, \delta, \eta, \zeta) \zeta^n = H_c(\alpha, \beta, \gamma, \delta, \eta, \zeta),
\end{equation}
where the coefficients $v_n$ satisfy a three term recurrence relation. This confluent Heun function $H_c(\zeta)$ must reduce to a polynomial, since otherwise it would increase exponentially as $\xi \to \infty$. $H_c(\zeta)$ reduces to a polynomial if two conditions are met.\cite{DowningHeun} Firstly, we need to adhere to
\begin{equation}
\label{eq24548}
	\tfrac{\delta}{\alpha} + \tfrac{1}{2} (\beta + \gamma) + N + 1 = 0, \quad N = 1, 2... 
\end{equation}
or, upon solving for energy and noting the restriction $|\varepsilon b | < 1$, we have
\begin{equation}
\label{eq29}
	\varepsilon_{N, m}^{QES} b = \pm \left\{ 1 - \left( \frac{\tfrac{R}{b} - m - \tfrac{1}{2}}{ N+ \tfrac{3}{2} - \tfrac{R}{b} + |m+1| } \right)^2 \right\}^{1/2},
\end{equation}
which ensures that the $(N+1)$th coefficient in the series expansion is a polynomial in $\eta$ of order $N + 1$. The second necessary condition is to find some value of $\delta$ that is a root of that polynomial, such that the coefficient $v_{N+1}$ is zero and hence (due to the recurrence relationships) all successive coefficients are also zero. Then the series has been truncated and $H_c(\zeta)$ is simply a confluent Heun polynomial. For clarity, in Eq.~\eqref{eq29} $N$ is the degree of the polynomial solution and is not related to the number of nodes of the full wave function.

Notably, Eq.~\eqref{eq24548} can only be satisfied for a restricted set of rotating, vortex states (defined by a certain quantum number $m$) within the interval $-\infty < m \le m^{*}$, where $m^{*} = m^{*}(R/b, N)$. Gradually decreasing the magnetic field strength (decreasing $R/b$) will lead to successive vortex states undergoing a confinement-deconfinement transition, until finally $m^{*} = -1$ and only negatively rotating states remain.

As example solutions of this model, let us set consider the $m=1$ state with $N = 1, 2, 3$ respectively. Upon solving for the roots of the resultant quadratic, cubic and quartic equations respectively in $\delta$ from the second condition, one finds the magnetic field strengths $R/b$ for each $N$ respectively. For $N=1$, we obtain the solution $R/b = 1.66$ which from Eq.~\eqref{eq29}, corresponds to the energy level  $\varepsilon_{1, 1}b = \pm 0.998$. Similarly for $N=2$, we find the solutions $R/b = 2.15, 1.66$ with associated energy levels $\varepsilon_{2, 1}b = \pm 0.981, \pm 0.999$ respectively. Finally, for $N=3$ we obtain the field strengths $R/b = 2.63, 2.14, 1.66$ which lead to the quantized energies $\varepsilon_{3, 1}b = \pm 0.956, \pm 0.989, \pm 0.999$ respectively.

\begin{figure}[btp]
 \includegraphics[width=0.45\textwidth]{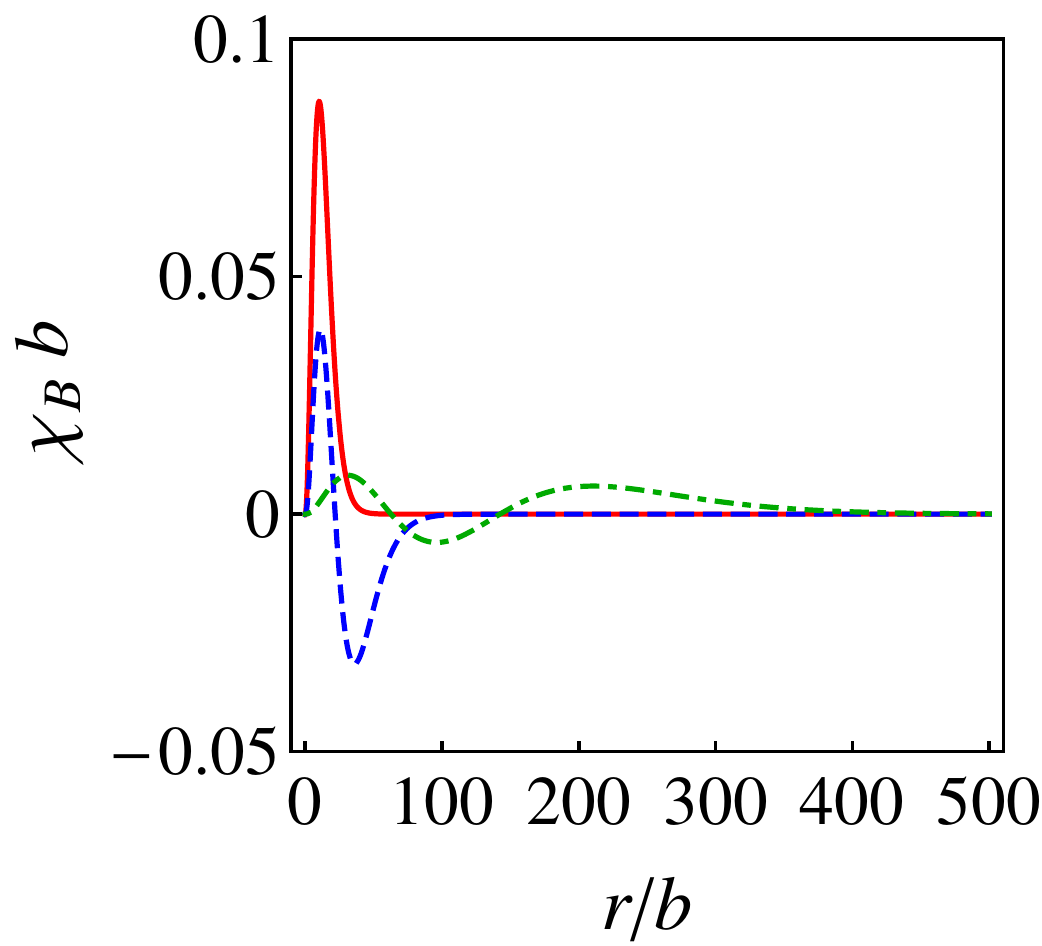}
 \caption{(Color online) Plots of the (non-normalized) radial wavefunction component $\chi_B$, through Eq.~\eqref{eq21}, for the $N=3$, $m=1$ QES results. Shown are a ground state (solid, red line) with $\varepsilon_{3, 1}b = 0.956, R/b = 2.63$, a first excited state (dashed, blue line) with $\varepsilon_{3, 1}b = 0.989, R/b = 2.14$, and a second excited state (dot-dashed, green line) with $\varepsilon_{3, 1}b = 0.999, R/b = 1.66$.}
 \label{fig3}
\end{figure}

We plot in Fig.~\ref{fig3} these aforementioned $N=3$ QES wavefunction components, through Eq.~\eqref{eq21}, which provide examples of states with an increasing number of nodes. We show a ground state (solid, red line), a first excited state (dashed, blue line) and a second excited state (dot-dashed, green line). As usual, higher node states are less tightly localized. The position of the energy level strongly influences the wavefunction decay through the key parameter $\kappa$, defined in Eq.~\eqref{eq22}, which appears in the exponent of $\chi_B$. Therefore, we have found that in a smooth, regularized model truly bound modes of 2D Dirac-Weyl excitations do exist and the excitations are limited to some maximum vorticity $m^{*}$ depending on the strength and spatial extent of the magnetic quantum dot.

For completeness, the zero-energy states of the system are simply obtained from Eq.~\eqref{eq123456} as
\begin{equation}
\label{299}
\left( \begin{array}{c}
\chi_A \\ \chi_B
 \end{array} \right) \sim  \left(
 \begin{array}{c}
 0 \\  r^{-(m+1)} \left( r + R \right)^{\tfrac{R}{b}} e^{- \tfrac{r}{b}}
 \end{array}
\right), \quad m \le -1.
\end{equation}
Again, these states possess the expected chiral property, and show the prohibition of non-negative $m$ vortex states. A similar analysis for a magnetic quantum ring problem is detailed in Appendix~\ref{sec5}, and again shows the characteristic behavior of excluding more and more positive angular momentum states from forming with decreasing magnetic field strength $R/b$ - an effect which seemingly does not have a parallel for electric potential wells with changing depth.

\section{\label{conc}Conclusions}

Confinement of 2D Dirac-Weyl particles in nonuniform magnetic fields, a problem of continued interest to the mesoscopic community, has been reconsidered. We have shown, with magnetic field profiles including examples of magnetic quantum dots and rings, how such traps should be of a long-range nature to hold bound states. We revealed how bound states with one sign of angular momentum are completely removed, in order to maintain a localized state, in the limit of a strongly inhomogeneous magnetic field. In reaching this limit, one will see a succession of confinement-deconfinement transitions as states disappear into the continuum once their value of angular momentum becomes prohibited. We hope experimental realization of such magnetic confinement can be achieved in the near future. 

Furthermore, in graphene certain configurations of strain can lead to pseudomagnetic fields,\cite{Guinea} which conserve time-reversal symmetry across the two valleys $K$ and $K'$. Thus our results suggest to achieve bound states mechanically one needs to create very delicate strain configurations, giving rise to long-range magnetic fields decaying like $1/r$. The sign of the effective magnetic field will be opposite in the two valleys, giving rise to states rotating only with positive angular momentum in one valley and only negative angular momentum in the other valley. This imbalance, which could become tunable after applying a real magnetic field, may be useful in future mesoscopic devices exploiting valley filtering.

\section*{Acknowledgments}

We acknowledge financial support from the CNRS and from the ANR under Grant No. ANR-14-CE26-0005 Q-MetaMat, as well as the EU H2020 RISE project CoExAN (Grant No. H2020-644076), EU FP7 ITN NOTEDEV (Grant No. FP7-607521), the FP7 IRSES projects CANTOR (Grant No. FP7-612285), QOCaN(Grant No. FP7-316432), and InterNoM (Grant No. FP7-612624). We would like to thank M.~Starr and L.~Foxx for fruitful discussions and S.~Zadinia for a critical reading of the manuscript.

\begin{appendix}

\section{\label{sec5}Dirac electron in a magnetic quantum ring}

It is straightforward to adapt the magnetic dot problem of Sec.~\ref{sec2} to describe a toy model of a magnetic quantum ring, defined by
\begin{equation}
\label{eq90}
 B_z (r) = \frac{\hbar}{e} \frac{1}{b r} \Theta(r-R),
\end{equation}
where $\Theta(z)$ is Heaviside's step function. Now, inside the ring $(r\le R)$ we have the usual free particle solution in terms of a Bessel function of the first kind $\chi_B^{I} = \tfrac{c_{I}}{b} J_{|m+1|}(\varepsilon r)$, and outside the ring $(r>R)$ we again have a wavefunction component like that in Eq.~\eqref{eq33}. Matching both wavefunction components at the ring boundary $r=R$ yields the following transcendental equation for the allowed eigenvalues
\begin{multline}
\label{eq91}
2 \kappa a_{II} \frac{U(a_{II}+1, b_{II}+1, \tfrac{2 \kappa R}{b})}{U(a_{II}, b_{II}, \tfrac{2 \kappa R}{b})} - \varepsilon b \frac{J_{|m+1|+1}(\varepsilon R)}{J_{|m+1|}(\varepsilon R)} \\
+ \kappa - 1 = 0.
\end{multline}
Solutions of Eq.~\eqref{eq91} show how, even in a true ring with an asymptotically decaying field, bound states exist. The dependence on the parameters of the system in Eq.~\eqref{eq91} is simple in the limit $R/b << 1$, when Eq.~\eqref{37b} is recovered. This result then forces one to introduce a caveat to the belief bound states do not arise in magnetic rings. The chirality of the ground state is illustrated via the zero-modes, with wavefunction components
\begin{equation}
\label{34388}
\left( \begin{array}{c}
\chi_B^I \\ \chi_B^{II}
 \end{array} \right) \sim r^{-(m+1)} \left(
 \begin{array}{c}
 1 \\ e^{\tfrac{R-r}{b}}
 \end{array}
\right), ~ \chi_A^I = \chi_A^{II} = 0, ~ m \le -1,
\end{equation}
showing the complete occupation on $B$ sites only.

\end{appendix}

\end{document}